\documentclass{emulateapj}
\usepackage{apjfonts}

\shorttitle{BLACK HOLE MASS IN ARP~151}
\shortauthors{BENTZ ET AL.}

\begin{document}

\title{First Results from the Lick AGN Monitoring Project: The Mass of
the Black Hole in Arp~151}

\author{ 
Misty~C.~Bentz\altaffilmark{1},
Jonelle~L.~Walsh\altaffilmark{1},
Aaron~J.~Barth\altaffilmark{1},
Nairn~Baliber\altaffilmark{2,3},
Nicola~Bennert\altaffilmark{2,4},
Gabriela~Canalizo\altaffilmark{4,5},
Alexei~V.~Filippenko\altaffilmark{6},
Mohan~Ganeshalingam\altaffilmark{6},
Elinor~L.~Gates\altaffilmark{7},
Jenny~E.~Greene\altaffilmark{8,9},
Marton~G.~Hidas\altaffilmark{2,3},
Kyle~D.~Hiner\altaffilmark{4,5},
Nicholas~Lee\altaffilmark{6},
Weidong~Li\altaffilmark{6},
Matthew~A.~Malkan\altaffilmark{10},
Takeo~Minezaki\altaffilmark{11},
Frank~J.~D.~Serduke\altaffilmark{6},
Joshua~H.~Shiode\altaffilmark{6},
Jeffrey~M.~Silverman\altaffilmark{6},
Thea~N.~Steele\altaffilmark{6},
Daniel~Stern\altaffilmark{12},
Rachel~A.~Street\altaffilmark{2,3},
Carol~E.~Thornton\altaffilmark{1},
Tommaso~Treu\altaffilmark{2,13},
Xiaofeng~Wang\altaffilmark{6,14},
Jong-Hak~Woo\altaffilmark{2,9,10}, and
Yuzuru~Yoshii\altaffilmark{11,15}
}

\altaffiltext{1}{Department of Physics and Astronomy,
                 4129 Frederick Reines Hall,
                 University of California,
                 Irvine, CA 92697;
                 mbentz@uci.edu } 

\altaffiltext{2}{Physics Department, 
                 University of California, 
                 Santa Barbara, CA 93106.}

\altaffiltext{3}{Las Cumbres Observatory Global Telescope, 
                  6740 Cortona Dr. Ste. 102, 
                  Goleta, CA 93117.}

\altaffiltext{4}{Institute of Geophysics and Planetary Physics,
                 University of California,
                 Riverside, CA 92521.}

\altaffiltext{5}{Department of Physics and Astronomy,
                 University of California,
                 Riverside, CA 92521.}

\altaffiltext{6}{Department of Astronomy, 
                 University of California,
                 Berkeley, CA 94720-3411.}

\altaffiltext{7}{Lick Observatory,
                 P.O. Box 85, 
                 Mount Hamilton, CA 95140.}

\altaffiltext{8}{Princeton University Observatory,
                 Princeton, NJ 08544.}

\altaffiltext{9}{Hubble Fellow.}

\altaffiltext{10}{Department of Physics and Astronomy, 
                 University of California, 
                 Los Angeles, CA 90024.}

\altaffiltext{11}{Institute of Astronomy, 
                 School of Science, University of Tokyo, 
                 2-21-1 Osawa, Mitaka, Tokyo 181-0015, Japan.}

\altaffiltext{12}{Jet Propulsion Laboratory, 
                  California Institute of Technology, 
                  MS 169-527, 4800 Oak Grove Drive, 
                  Pasadena, CA 91109.}

\altaffiltext{13}{Sloan Fellow, Packard Fellow.}

\altaffiltext{14}{Physics Department and Tsinghua 
                  Center for Astrophysics (THCA), Tsinghua
                  University, Beijing, 100084, China.}

\altaffiltext{15}{Research Center for the Early Universe, 
                  School of Science,
                  University of Tokyo, 7-3-1 Hongo, Bunkyo-ku, 
                  Tokyo 113-0033, Japan.}

\begin{abstract}

We have recently completed a 64-night spectroscopic monitoring
campaign at the Lick Observatory 3-m Shane telescope with the aim of
measuring the masses of the black holes in 13 nearby ($z < 0.05$)
Seyfert~1 galaxies with expected masses in the range $\sim 10^6 -
10^7$~M$_{\odot}$. We present here the first results from this project
-- the mass of the central black hole in Arp~151.  Strong variability
throughout the campaign led to an exceptionally clean H$\beta$ lag
measurement in this object of $4.25^{+0.68}_{-0.66}$~days in the
observed frame.  Coupled with the width of the H$\beta$ emission line
in the variable spectrum, we determine a black hole mass of $(7.1 \pm
1.2) \times 10^6$~M$_{\odot}$, assuming the \citeauthor{onken04}\
normalization for reverberation-based virial masses.  We also find
velocity-resolved lag information within the H$\beta$ emission line
which clearly shows infalling gas in the H$\beta$-emitting region.
Further detailed analysis may lead to a full model of the geometry and
kinematics of broad line region gas around the central black hole in
Arp~151.

\end{abstract}


%%%%%%%%%%%%%%%%%%%%%%%%%%%
%%%%%%%%%%%%%%%%%%%%%%%%%%%
\section{Introduction}
%%%%%%%%%%%%%%%%%%%%%%%%%%%
%%%%%%%%%%%%%%%%%%%%%%%%%%%

Reverberation mapping (\citealt{blandford82,peterson93}) is the most
successful method employed for measuring the central black hole mass
in Type~1 active galactic nuclei (AGNs).  Rather than relying on
spatially-resolved observations, reverberation mapping resolves the
influence of the black hole in the time domain through spectroscopic
monitoring of changes in the continuum flux and the delayed response,
or ``echo,'' in the broad emission lines.  The time lag between these
changes, $\tau$, depends on the light-travel time across the
broad-line region (BLR).  Combining the radius of the BLR, $c \tau$,
with the velocity width of the broad emission line gives the virial
mass of the central black hole.

To date, successful reverberation-mapping studies have been carried
out for approximately 36 active galaxies (compiled by
\citealt{peterson04,peterson05}), mostly probing black hole masses in
the range $10^7 - 10^9$~M$_{\odot}$.  Studies of lower-mass AGNs have
been restricted by their lower luminosities, requiring telescopes
larger than the typical 1.5-m apertures that have been employed.  With
the goal of extending the mass range probed by reverberation studies,
we have carried out a 64-night spectroscopic monitoring campaign on
the Lick Observatory 3-m Shane telescope, targeting 13 AGNs with
expected black hole masses in the range $\sim 10^6 -
10^7$~M$_{\odot}$.  We present here the first results from this
project: an analysis of the H$\beta$ reverberation in the nearby
($z=0.0211$) Seyfert galaxy Arp~151 (Mrk~40).  Full campaign details
and results will be presented in a series of forthcoming papers.

%%%%%%%%%%%%%%%%%%%%%%%%%%%
%%%%%%%%%%%%%%%%%%%%%%%%%%%
\section{Observations}
%%%%%%%%%%%%%%%%%%%%%%%%%%%
%%%%%%%%%%%%%%%%%%%%%%%%%%%

%%%%%%%%%%%%%%%%%%%%%%%%%%%
\subsection{Photometry}
%%%%%%%%%%%%%%%%%%%%%%%%%%%

Broad-band Johnson $B$ images of Arp~151 were obtained at the 32-inch
Tenagra II telescope in Southern Arizona most nights between calendar
dates 2008 February 26 and May 15.  Typical exposure times were $2
\times 300$\,s.

The images were reduced following standard techniques. The flux of the
AGN was measured through a circular aperture of radius 4.35\arcsec,
and differential photometry was obtained relative to 8 stars within
the field.\footnote{A simple model of the host galaxy surface
brightness profile from the ground-based images indicates that $\sim
40$\% of the light within this aperture comes from the host galaxy
starlight.}  Absolute flux calibrations were determined on a
photometric night using the Landolt SA-101 and SA-109 standard star
fields.  The calibrated light curve is shown in Figure~1.  For the
cross-correlation analysis, the $B$-band magnitudes were converted to
fluxes.

\begin{deluxetable}{lcccc}
\tablecolumns{5}
\tablewidth{230pt}
\tablecaption{Light-Curve Statistics}
\tablehead{
\colhead{Time Series} &
\colhead{$N$} &
\colhead{$T_{\rm median}$} &
\colhead{$\langle f \rangle$\tablenotemark{a}} &
\colhead{$\langle \sigma_{\rm f}/f \rangle$} \\
\colhead{(1)} &
\colhead{(2)} &
\colhead{(3)} &
\colhead{(4)} &
\colhead{(5)}}
\startdata

$B$ band & 60 & 1.03 & $1.46 \pm 0.04$  & 0.0197  \\
5100 \AA & 43 & 1.02 & $1.58 \pm 0.18$  & 0.0614  \\
H$\beta$ & 43 & 1.02 & $1.14 \pm 0.19$  & 0.0241  \\
\enddata 

\tablecomments{Columns are presented as follows: (1) feature; (2)
               number of observations; (3) median
               sampling rate in days; (4) mean flux and
               standard deviation; and (5) mean fractional error.}

\tablenotetext{a}{Flux densities are in units of
                  $10^{-15}$~ergs~s$^{-1}$~cm$^{-2}$~\AA$^{-1}$;
                  emission-line fluxes are in units of
                  $10^{-13}$~ergs~s$^{-1}$~cm$^{-2}$.  }

\end{deluxetable}

\begin{figure}
\epsscale{1.235}
\plotone{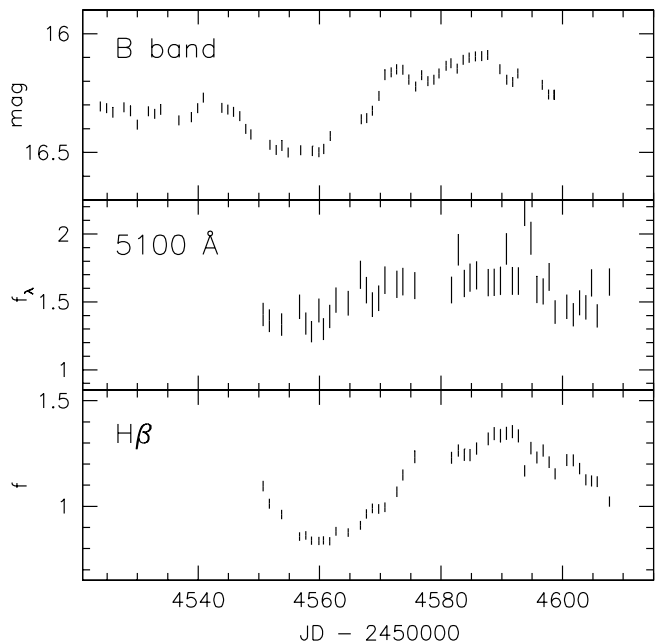}
\caption{Light curves for Arp~151.  From top to bottom, the panels
         are: $B$-band magnitude, observed continuum flux density at
         rest-frame 5100\,\AA, and H$\beta$ emission-line flux.  The
         units are as given in Table~1 for 5100\,\AA\ and H$\beta$.}
\end{figure}

%%%%%%%%%%%%%%%%%%%%%%%%%%%
\subsection{Spectroscopy}
%%%%%%%%%%%%%%%%%%%%%%%%%%%

Spectroscopic monitoring was carried out at the Lick Observatory 3-m
Shane telescope with the Kast dual spectrograph.  Arp~151 was observed
from March 24 -- May 20 on a total of 43 nights.  We restricted our
observations to the Kast red-side CCD and employed the
600~lines~mm$^{-1}$ grating with spectral coverage over the range
4300--7100\,\AA.  Spectra were obtained at a fixed position angle of
90\degr\ through a 4\arcsec-wide slit.  Exposures were usually $2
\times 600$\,s at an airmass of $\sim 1.1$, resulting in a typical
signal-to-noise ratio ($S/N$) of $\sim 100$ per pixel at rest-frame
5100\,\AA\ in the combined spectra.  The images were reduced with IRAF
and the spectra were extracted with a width of 13 pixels
(10.1\arcsec), with sky regions of width 6 pixels beginning at a
distance of 19 pixels.  Flux calibrations were determined from nightly
spectra of standard stars.

To mitigate the effects of slit losses and variable seeing and
transparency, we employed the spectral scaling algorithm of
\citet{vangroningen92} to scale the total flux of the narrow
[\ion{O}{3}] $\lambda \lambda 4959, 5007$ doublet in each individual
spectrum to match the [\ion{O}{3}] flux in a reference spectrum
created from the spectra obtained on the 10 nights with the best
weather and seeing conditions.  This method accounts for differences
in the overall flux scale, as well as small wavelength shifts and
small differences in spectral resolution, and has been shown to result
in spectrophotometric accuracies of $\sim 2$\% \citep{peterson98a}.

Finally, the spectroscopic light curves were measured.  The continuum
flux was measured in the observed-frame window 5187--5227\,\AA\
(rest-frame 5100\,\AA).  The flux in this window is a combination of
the AGN continuum flux, which is variable, and a constant component
from the host-galaxy starlight.  The H$\beta$ flux was measured by
fitting a linear continuum under the H$\beta$ emission line,
determined from the continuum flux at 4850--4900\,\AA\ and
5175--5225\,\AA, and then integrating the emission-line flux above the
continuum from 4900--5050\,\AA.  This technique includes the flux
contribution from the narrow H$\beta$ emission line, which is simply a
constant offset.  The resultant light curves are presented in
Figure~1, with statistical properties listed in Table~1.  Figure~2
shows the mean and root-mean-square (rms) spectra, where the rms
spectrum shows the standard deviation of all the spectra relative to
the mean spectrum (i.e., the variable components of the spectra).

\begin{figure}
\epsscale{1.235}
\plotone{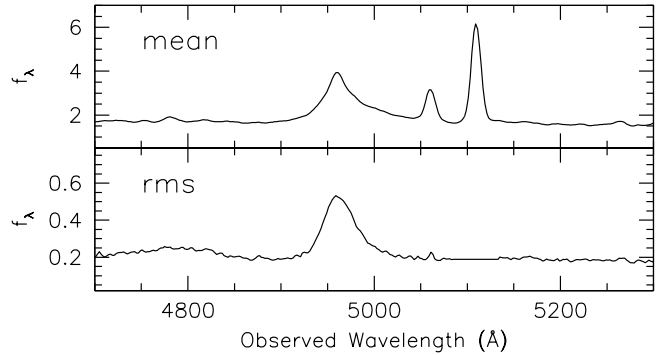}
\caption{Mean and variable (rms) spectra of Arp~151 in the H$\beta$
         region.  The narrow lines have been removed from the rms
         spectrum.}
\end{figure}

%%%%%%%%%%%%%%%%%%%%%%%%%%%
%%%%%%%%%%%%%%%%%%%%%%%%%%%
\section{Analysis}
%%%%%%%%%%%%%%%%%%%%%%%%%%%
%%%%%%%%%%%%%%%%%%%%%%%%%%%
\subsection{Time-Series Analysis}
%%%%%%%%%%%%%%%%%%%%%%%%%%%

For the time-series analysis, we place more emphasis on the $B$-band
light curve as the driving, continuum light curve, although we also
consider the lower $S/N$ light curve of the 5100\,\AA\ flux.  To
determine the average time lag between variations in the continuum
flux and variations in the H$\beta$ emission-line flux, we follow the
standard practice of cross-correlating the light curves using the
interpolation cross-correlation function (ICCF) method
(\citealt{gaskell86,gaskell87}) as well as the discrete correlation
function (DCF) method \citep{edelson88}, with the \citet{white94}
modifications to both.  The resultant cross-correlation functions are
shown in Figure~3. The uncertainties in the time lag are determined
using the Monte Carlo ``flux randomization/random subset sampling''
method described by \citet{peterson98b,peterson04}.  In short, the
method samples a random subset of the data points in the light curves,
randomizes the fluxes by applying a Gaussian deviation within the flux
uncertainties, and cross-correlates the modified light curves.  The
procedure is carried out 1000 times, and a distribution of lag
measurements is built up.  We include two specific measurements of the
lag in Table~2: $\tau_{\rm peak}$, the location of the maximum of the
cross-correlation function $r_{\rm max}$; and $\tau_{\rm cent}$, the
centroid of the points near the peak of the function with $r \geq
0.8r_{\rm max}$.  The uncertainties on $\tau_{\rm peak}$ and
$\tau_{\rm cent}$ are set such that 15.87\% of the Monte Carlo
realizations fall below the range indicated by the uncertainties, and
15.87\% fall above this range (i.e., $1 \sigma$ uncertainties for a
Gaussian distribution).  We measure an average observed-frame lag of
$\tau_{\rm cent} = 4.25^{+0.68}_{-0.66}$\,days between the $B$-band
and H$\beta$ light curves.  The values listed in Table~2 are corrected
for time dilation effects.

\begin{figure}
\epsscale{1.235}
\plotone{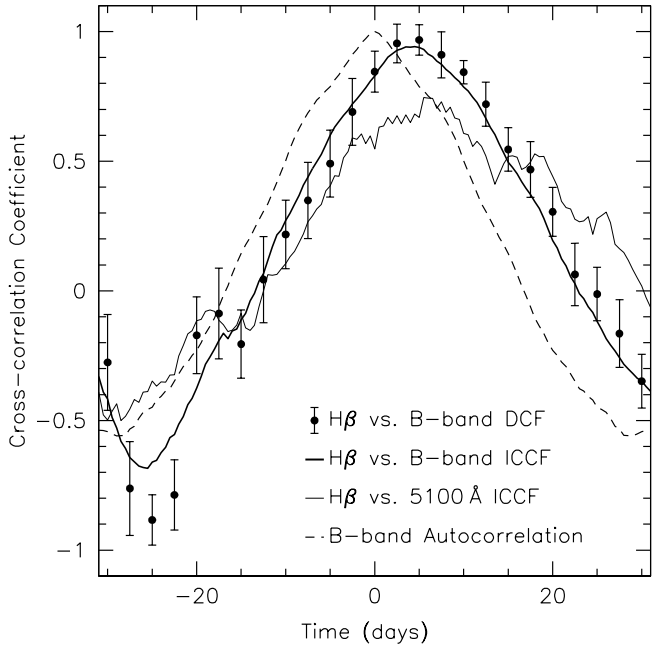}
\caption{Cross-correlation functions for Arp~151.  The heavy line is
         the cross-correlation of the $B$-band flux and H$\beta$
         emission, and the data points show the discrete correlation
         between the same.  The cross-correlation of the 5100\,\AA\
         flux and H$\beta$ (thin solid line), while noisier, is
         consistent.  The dashed line is the $B$-band auto-correlation
         function, which peaks at zero lag as expected.}
\end{figure}

%%%%%%%%%%%%%%%%%%%%%%%%%%%
\subsection{Line Width Measurement}
%%%%%%%%%%%%%%%%%%%%%%%%%%%

The width of the broad H$\beta$ emission line was measured in the mean
and rms spectra.  We report here two separate measures of the line
width: the full-width at half-maximum flux (FWHM) and the line
dispersion, $\sigma_{\rm line}$, which is the second moment of the
emission-line profile \citep{peterson04}.  The uncertainties in the
line widths are again set using Monte Carlo random subset sampling
methods.  In this case, a random subset of the spectra is chosen and a
mean and rms spectrum are created, from which the FWHM and
$\sigma_{\rm line}$ are measured.  A distribution of line-width
measurements is built up through 1000 realizations, from which we take
the mean and the standard deviation to be the line width and its
typical uncertainty, respectively.  Additional systematic errors, such
as those due to the exact determination of the continuum contribution,
are not included in these estimates of the uncertainty.  The line
widths presented in Table~2 have been corrected for the resolution of
the spectrograph following \citet{peterson04}.

%%%%%%%%%%%%%%%%%%%%%%%%%%%
\subsection{Black Hole Mass}
%%%%%%%%%%%%%%%%%%%%%%%%%%%

Following the usual assumption that the BLR kinematics are
gravitationally driven, the black hole mass is determined via the
virial equation

\begin{equation}
    M_{\rm BH} = f \frac{c \tau v^2}{G},
\end{equation}

\noindent where $\tau$ is the mean time delay for the region of
interest (here, the H$\beta$-emitting region), $v$ is the velocity of
gas in that region, $c$ is the speed of light, $G$ is the
gravitational constant, and $f$ is a scaling factor of order unity
that depends on the detailed geometry and kinematics of the region.

With the premise that the $M_{\rm BH} - \sigma_{\star}$ relationship
for local, quiescent galaxies holds for AGNs and their host galaxies,
\citet{onken04} find that $\langle f \rangle \approx 5.5$ for
reverberation-based masses.  This particular scaling is appropriate
when $\tau_{\rm cent}$ and $\sigma_{\rm line,rms}$ are used for the
lag and line width in the black hole mass determination.  For the
measurements presented here, the \citeauthor{onken04}\ normalization
gives $M_{\rm BH} = (7.1 \pm 1.2) \times 10^6$~M$_{\odot}$.
Individual reverberation masses, however, are subject to a typical
factor of $2-3$ uncertainty (\citeauthor{onken04}), likely due to
differences in the intrinsic, but unknown, $f$ value for each
individual system.  Also listed in Table~2 is the ``virial product,''
assuming $f = 1$.  The black hole mass in Arp~151 is smaller than the
estimate based on the stellar velocity dispersion ($\sigma_{\star} =
124 \pm 12$~km~s$^{-1}$; \citealt{greene06}) and the $M_{\rm BH} -
\sigma_{\star}$ relationship of \citet{tremaine02}, which predicts
$2.0^{+1.7}_{-1.0} \times 10^7$~M$_{\odot}$, but the two are
consistent within the known scatter for reverberation-based masses and
the $M_{\rm BH} - \sigma_{\star}$ relationship.

\begin{figure}
\epsscale{1.235}
\plotone{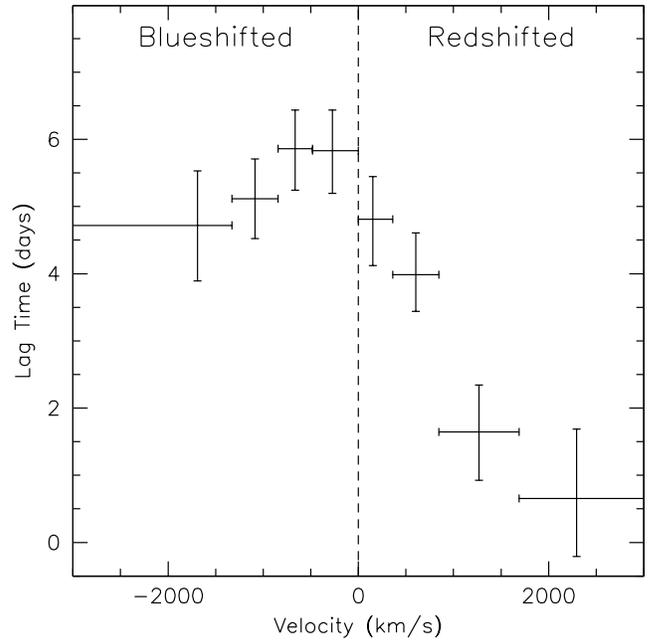}
\caption{Velocity-resolved H$\beta$ lag measurements for Arp~151.  The
         error bars in the lag direction are the $1\sigma$
         uncertainties on the lag measurements, while in the velocity
         direction they show the size of the velocity bin (which
         extend to the line integration limits on both sides).  A
         clear signature of blueshifted gas with long lags and
         redshifted gas with short lags is detected, indicating radial
         infall of BLR gas.}
\end{figure}

%%%%%%%%%%%%%%%%%%%%%%%%%%%
%%%%%%%%%%%%%%%%%%%%%%%%%%%
\section{Velocity-Resolved Time Delay Measurements}
%%%%%%%%%%%%%%%%%%%%%%%%%%%

\begin{deluxetable}{lcc}
\tablecolumns{3}
\tablewidth{230pt}
\tablecaption{Rest-Frame Reverberation Measurements}
\tablehead{
\colhead{Measurement} &
\colhead{Value} &
\colhead{Units}} 
\startdata

$\tau_{\rm cent}$           & $4.17^{+0.67}_{-0.64}$  & days        \\
$\tau_{\rm peak}$           & $3.67^{+0.74}_{-0.24}$  & days        \\
$\sigma_{\rm line}$~(mean)  & $1738 \pm 10$           & km~s$^{-1}$ \\
FWHM~(mean)                 & $2842 \pm 59$           & km~s$^{-1}$ \\
$\sigma_{\rm line}$~(rms)   & $1261 \pm 37$           & km~s$^{-1}$ \\
FWHM~(rms)                  & $2283 \pm 143$          & km~s$^{-1}$ \\
$c \tau v^2 / G $           & $1.3 \pm 0.2$    & $10^6$~M$_{\odot}$ \\
$M_{\rm BH}$                & $7.1 \pm 1.2$    & $10^6$~M$_{\odot}$ \\

\enddata
\end{deluxetable}

A key goal of reverberation mapping is to recover the full transfer
function (time delay vs. velocity structure) responsible for the shape
of the emission-line light curve in response to the driving continuum
light curve. Determining the transfer function is the most promising
method to potentially provide detailed information on the geometry and
kinematics of the BLR.  While some hints of the transfer function
shape have been seen in certain high-quality reverberation data sets
(i.e., \citealt{horne91}), a full recovery of the transfer function
has not yet been achieved.

We carried out an initial analysis of the velocity-resolved time lag
information for Arp~151 by binning the H$\beta$ emission line in
velocity space, where each of the eight bins contained an equal amount
of variable flux.  Eight light curves were created, one for each
velocity bin, and the light curves were each cross-correlated with the
$B$-band light curve using the methods in \S3.1.  Figure~4 shows the
results of this analysis: there is a clear gradient in the gas
response where the blueshifted H$\beta$ emission lags the response in
the redshifted H$\beta$ emission.  This is the typical signature of
radial infall: the gas on the far side of the AGN is moving toward us,
and the gas on the near side is moving away from us.  Outflowing
(i.e., wind-driven) gas would produce the opposite effect (short lags
blueshifted and long lags redshifted) while pure rotation would
produce a symmetric pattern around zero velocity.  Only outflow
specifically precludes a determination of $M_{\rm BH}$ due to the
non-gravitational motion of the BLR gas.  While the kinematics of the
BLR H$\beta$-emitting gas in Arp~151 show a strong signature of radial
infall, a full two-dimensional echo map of the velocity and time delay
structure in the spectra (e.g., \citealt{welsh91}) must be recovered
before we can fully explain the details of the BLR geometry and
kinematics.  Such an effort is beyond the scope of this paper, but is
currently being pursued.

The results presented here demonstrate the clearest signature of
gravitational infall in the BLR of an AGN to date.  Some indications
of infalling gas have been seen in the \ion{C}{4} broad line response
in other objects, such as NGC~5548 (e.g., \citealt{crenshaw90,done96})
and Fairall~9 \citep{koratkar89}.  As such, AGN BLRs seem to commonly
exhibit signatures of infalling gas (but see
\citealt{maoz93,kollatschny03}).  The $f$ value in Equation~1 is
directly dependent on the kinematics of the BLR, thereby resulting in
a different $f$ value for an AGN BLR with radial infall than for a BLR
with Keplerian rotation.  The \citet{onken04} value of $\langle f
\rangle \approx 5.5$ is determined empirically and is independent of
specific BLR models.  Subsumed into the population average, $\langle f
\rangle$, are the signatures of kinematic and geometric states that
are common among AGNs with reverberation results.  While it is almost
certain that the individual details of each BLR will result in
somewhat different intrinsic $f$ values for every AGN (e.g.,
\citealt{collin06}), there is currently no reason to expect that the
$f$ value for Arp~151 is wildly discrepant from the population
average, or to suspect the mass derived here of having uncertainties
larger than those typically expected for reverberation masses.

%%%%%%%%%%%%%%%%%%%%%%%%%%%
%%%%%%%%%%%%%%%%%%%%%%%%%%%
\section{Summary}
%%%%%%%%%%%%%%%%%%%%%%%%%%%
%%%%%%%%%%%%%%%%%%%%%%%%%%%

We have presented the first light curves and reverberation analysis
from our AGN monitoring campaign at Lick Observatory.  We detect a
clear lag in the broad H$\beta$ emission-line response to changes in
the continuum flux for Arp~151, and we present a measurement of the
black hole mass assuming the \citet{onken04} normalization.  Initial
analysis of velocity-resolved time delays in the H$\beta$ line shows a
strong signature of infalling gas, but further work is needed to map
out the detailed structure and kinematics of the BLR in Arp~151.

We see strong variability in other emission lines, including
H$\alpha$, H$\gamma$, and \ion{He}{2}, the analysis of which will be
included in future papers.  In addition, we have a {\it Hubble Space
Telescope} Cycle 17 program to image the host galaxies of the AGNs in
this sample, allowing us to correct their spectroscopic luminosities
for starlight and apply these new results to the low end of the
radius--luminosity relationship for AGNs \citep{bentz06a,bentz08b},
which is the primary calibration for all single-epoch mass estimates
for broad-lined AGNs.

\acknowledgements

We thank the Lick Observatory staff for their tireless support during
this project.  We also thank Brad Peterson for helpful conversations
and the use of his analysis software.  This work was supported by NSF
grants AST--0548198 (UC Irvine), AST--0607485 (UC Berkeley),
AST--0642621 (UC Santa Barbara), and AST--0507450 (UC Riverside), as
well as the TABASGO Foundation (UC Berkeley).

%\bibliographystyle{apj}
%\bibliography{mbentz}

\begin{thebibliography}{24}
\expandafter\ifx\csname natexlab\endcsname\relax\def\natexlab#1{#1}\fi

\bibitem[{{Bentz} {et~al.}(2006){Bentz}, {Peterson}, {Pogge}, {Vestergaard}, \&
  {Onken}}]{bentz06a}
{Bentz}, M.~C., {et~al.} 2006, \apj, 644, 133

\bibitem[{{Bentz} {et~al.}(2008)}]{bentz08b}
---. 2008, \apj, {submitted}

\bibitem[{{Blandford} \& {McKee}(1982)}]{blandford82}
{Blandford}, R.~D., \& {McKee}, C.~F. 1982, \apj, 255, 419

\bibitem[{{Collin} {et~al.}(2006)}]{collin06}
{Collin}, S., {et~al.} 2006, \aap, 456, 75

\bibitem[{{Crenshaw} \& {Blackwell}(1990)}]{crenshaw90}
{Crenshaw}, D.~M., \& {Blackwell}, Jr., J.~H. 1990, \apjl, 358, L37

\bibitem[{{Done} \& {Krolik}(1996)}]{done96}
{Done}, C. \& {Krolik}, J.~H.  1996, \apj, 463, 144

\bibitem[{{Edelson} \& {Krolik}(1988)}]{edelson88}
{Edelson}, R.~A., \& {Krolik}, J.~H. 1988, \apj, 333, 646

%\bibitem[{{Gaskell} \& {Goosmann}(2008)}]{gaskell08}
%{Gaskell}, C.~M., \& {Goosmann}, R.~W. 2008, \apj, {submitted
%  (astro-ph/0805.4258)}

\bibitem[{{Gaskell} \& {Peterson}(1987)}]{gaskell87}
{Gaskell}, C.~M., \& {Peterson}, B.~M. 1987, \apjs, 65, 1

\bibitem[{{Gaskell} \& {Sparke}(1986)}]{gaskell86}
{Gaskell}, C.~M., \& {Sparke}, L.~S. 1986, \apj, 305, 175

\bibitem[{{Greene} \& {Ho}(2006)}]{greene06}
{Greene}, J.~E., \& {Ho}, L.~C. 2006, \apjl, 641, L21

\bibitem[{{Horne} {et~al.}(1991){Horne}, {Welsh}, \& {Peterson}}]{horne91}
{Horne}, K., {Welsh}, W.~F., \& {Peterson}, B.~M. 1991, \apjl, 367, L5

\bibitem[{{Kollatschny}(2003)}]{kollatschny03}
{Kollatschny}, W. 2003, \aap, 407, 461

\bibitem[{{Koratkar} \& {Gaskell}(1989)}]{koratkar89}
{Koratkar}, A.~P., \& {Gaskell}, C.~M. 1989, \apj, 345, 637

\bibitem[{{Maoz} {et~al.}(1993)}]{maoz93}
{Maoz}, D., {et~al.} 1993, \apj, 367, 493

\bibitem[{{Onken} {et~al.}(2004)}]{onken04}
{Onken}, C.~A., {et~al.} 2004, \apj, 615, 645

\bibitem[{{Peterson}(1993)}]{peterson93}
{Peterson}, B.~M. 1993, \pasp, 105, 247

\bibitem[{{Peterson} {et~al.}(1998{\natexlab{a}})}]{peterson98a}
{Peterson}, B.~M., {et~al.} 1998{\natexlab{a}}, \apj, 501, 82

\bibitem[{{Peterson} {et~al.}(1998{\natexlab{b}})}]{peterson98b}
---. 1998{\natexlab{b}}, \pasp, 110, 660

\bibitem[{{Peterson} {et~al.}(2004)}]{peterson04}
---. 2004, \apj, 613, 682

\bibitem[{{Peterson} {et~al.}(2005)}]{peterson05}
---. 2005, \apj, 632, 799

\bibitem[{{Tremaine} {et~al.}(2002)}]{tremaine02}
{Tremaine}, S., {et~al.} 2002, \apj, 574, 740

\bibitem[{{van Groningen} \& {Wanders}(1992)}]{vangroningen92}
{van Groningen}, E., \& {Wanders}, I. 1992, \pasp, 104, 700

\bibitem[{{Welsh} \& {Horne}(1991)}]{welsh91}
{Welsh}, W.~F., \& {Horne}, K. 1991, \apj, 379, 586

\bibitem[{{White} \& {Peterson}(1994)}]{white94}
{White}, R.~J., \& {Peterson}, B.~M. 1994, \pasp, 106, 879


\end{thebibliography}

\end{document}